# Researchers in an Entropy Wonderland: A Review of the Entropy Concept

Marko Popovic

Department of Chemistry and Biochemistry, Brigham Young University, Provo, UT 84602

> *"I intentionally chose the word Entropy as similar as possible to the word Energy…*
> *The energy of the universe is constant; the entropy of the universe tends to a maximum."*
> *Rudolf Clausius, 1867*

**Abstract:** *Entropy concept was introduced by Clausius 160 years ago, and has been continually enriched, developed and interpreted by the researchers in many different scientific disciplines ever since. Thermodynamics and other scientific disciplines face several simple but crucial questions concerning the entropy concept. Thus, it is frequently possible to notice a misuse of entropy. Sometimes unbelievable confusion in the literature is summarized by von Neumann's sentence: "No one knows what entropy really is." Four major questions stand before thermodynamics. (1) How many kinds of entropy are there? (2) What is the physical meaning of entropy? (3) Is entropy a subjective or an objective property? (4) Is entropy in any way related to information? This paper attempts to describe the roots, the conceptual history of this important concept and to describe the path of its development and application in different scientific disciplines. Through this we attempt to give some possible answers to the questions mentioned above.*

1. Introduction

Entropy concept is frequently used in many scientific disciplines: physics [1-70], equilibrium and non-equilibrium thermodynamics [1-25, 60-70], statistical mechanics [89,91,95], cosmology [50-59], life sciences [42-46, 86], chemistry and biochemistry [60-67], geosciences [85], linguistics [84], social sciences [83], and information theory [41,69,74, 88].

Before we proceed to review the entropy concept, it should be stressed that the concept is often used, misused and even abused in the literature [30, 79-82, 108]. The widespread use of entropy in many disciplines lead to many contradictions and misconceptions involving entropy, summarized in von Neumann's words: "*Whoever uses the term 'entropy' in a discussion always wins since no one knows what entropy really is, so in a debate one always has the advantage*" [80,81]. There are two main reasons why entropy concept is often used in different disciplines for presenting new theories [72]. First, no one knows what entropy really is [72, 80, 81], so no one will dare to challenge the new theory. Second, the word "entropy" is intriguing, mysterious, abstract, intellectual and fashionable, so it will attract attention of many to a new theory [72]. Kostic [30] reported that the misuse of entropy concept leads to "unbelievable confusion". The following sentence illustrates the confusion: "*The mathematical definition of Shannon entropy has the same form as entropy used in thermodynamics, and they share the same conceptual root in the sense that both measure the amount of randomness*" [72]. As will be sown below, this sentence is just partly valid for residual entropy $S_0$ only, and not valid for thermal entropy $S_{Therm}$. Residual entropy is not the first nor the only kind of entropy used in thermodynamics. Residual and Shannon entropy indeed share very similar definition, normal distribution and statistical coin model but constants in their equations ($k_B$ and $K$, respectively) are not the same [14]. However, thermal entropy $S_{Therm}$ represents a completely different kind of entropy. It uses Boltzmann distribution and Boltzmann constant $k_B$. Thus, such a sentence leads to confusion especially for readers not fluent in thermodynamics. From the perspective of particles that make a system, thermal entropy has been defined in several ways, some more correct than others. One of the definitions of thermal entropy is: "(thermal) *Entropy is a measure of the number of specific realizations or microstates*" [31]. Another definition states that entropy is: "*the configuration or arrangement of the atoms and molecules in the system*" [39]. Notice that opposite to [39] the above definition [31] uses more appropriate words: "*measure of the number of specific realizations or microstates*", instead of: "*configuration or arrangement*" to define thermal entropy. "Configuration and arrangement" are words that could be more appropriate related to residual or Shannon entropy. Another definition is the chemical thermodynamic definition: "*Qualitatively, entropy is simply a measure of how much the energy of atoms and molecules become more spread out in a process*" [32]. This definition is in line with Clausius' original intention to relate *Entropy* and *Energy*. Thermal entropy is according to Clausius [1-3] a part of the total energy content and measures how much of the total energy becomes useless. Thus, entropy represents a fraction of energy that cannot be extracted from the system and converted into the mechanical work.

Moreover, Schrödinger introduced "negentropy" as entropy taken with a negative sign ($S<0$) [6]. This demonstrates very well how deep the misunderstanding of the physical nature of entropy can be. Indeed, entropy is, just as energy, a nonnegative property according to the third law of thermodynamics [12, 90, 91, 94] and Clausius' [1-3] definition. Furthermore, the definition [30] that entropy is "*associated with energy*" is not very helpful. The phrases "*associated with*" and "*related to*" do not imply the physical meaning of entropy, nor the real nature of the relationship of entropy and energy.

A significant part of scientific community considers entropy as a subjective property [67,69,74]. Others insist that entropy is an objective property [70,76-78]. So, von Neumann was right – no one knows what entropy really is (subjective or objective, energy or something else, arrangement of particles or realization of microstates, negentropy, many kinds of entropy…). The following Neumann-Shannon anecdote expresses the frustration over two different quantities being given the same name. Shannon and his wife had a new baby. Von Neumann suggested them to name their son after the son of Clausius "entropy". Shannon decides to do so, to find out, in the years to follow, that people continually confuse his son with Clausius' son and also misuse and abuse the name [79].

The importance of the entropy concept is probably illustrated in a best manner by Sir Edington's words: "*The law that entropy always increases holds, I think, the supreme position among the laws of Nature…*" [4]. Einstein paid tribute to entropy: "*Classical thermodynamics is the only physical theory of universal content which I am convinced will never be overthrown, within the framework of applicability of its basic concepts*" [103].

This paper addresses four main questions: (1) How many kinds of entropy are there? (2) What is the physical meaning of entropy? (3) Is entropy a subjective or an objective property? (4) Is entropy in any way related to information? Answering these questions would give us a better insight into the entropy concept and its appropriate application.

## 2. Roots and genesis of the entropy concept

The term entropy is used by many authors in many scientific disciplines, resulting in many definitions and causing confusion about its meaning. Clausius created the term entropy as an extensive classical thermodynamic variable, a state function shown to be useful in characterizing the Carnot cycle. Two kinds of entropy; thermodynamic and Shannon entropy are commonly encountered in the literature. The total thermodynamic entropy includes residual entropy near zero kelvins and thermal entropy at temperatures above absolute zero [117]. Shannon entropy is A measure of uncertainty of arrangement of material carriers of information in a string [34]. This section reviews entropy used in macroscopic thermodynamics, statistical mechanics, information theory, as well as its applications in life sciences.

### 2.1. Clausius and Macroscopic Entropy

The story begins with recognition of low efficiency of steam engines used by the industry in the early and mid-19$^{th}$ century, an important period for thermodynamics. It saw the establishment of basic concepts such as thermodynamic system, its state and surroundings, state variables, equilibrium, and reversible and irreversible thermodynamic process. The machine inefficiency issue was attracting attention of many researchers in the mid-19$^{th}$ century, including a young physicist named Rudolf Clausius. In the early 1850s, Clausius set forth the concept of thermodynamic system, defining it as the material content of a macroscopic volume in space – the rest of the universe being its surroundings, and dividing it into three main categories: isolated, closed and open systems. A thermodynamic system is in a state defined through state variables, such as the amount of substance, volume, temperature, entropy, enthalpy, internal energy and pressure. A class of states of particular importance to thermodynamics is thermodynamic equilibrium, a state where there is no macroscopic flow of energy or matter between systems that are in equilibrium, or in case of an internal equilibrium between parts of a single system. All three kinds of thermodynamic systems can be in or out of state of thermodynamic equilibrium [116]. Thermodynamic states define thermodynamic processes: any change of state of a thermodynamic system represents a thermodynamic process [1-3]. Thermodynamic processes can be categorized in two ways. First, processes are categorized by state variables they affect, as: isothermal, isochoric, isobaric, isenthalpic, isentropic and adiabatic. Second, processes are categorized based on whether system and its surroundings are in or out of equilibrium during the process, resulting in two groups: reversible and irreversible processes. In a reversible process, the system and its surroundings are in a state of equilibrium during the entire process [35]. Equilibrium implies that the system and its surroundings have practically equal state variables, such as pressure or temperature, differing only by an infinitesimal amount that drives the process. The infinitesimal driving force enables the process to be reversed by an almost negligible – also infinitesimal change in some property of the system or the surroundings. Thus, the name: reversible process. A reversible process requires constant establishment of equilibrium between system and surroundings [35]. Because true equilibrium is a limiting state that takes an infinite amount of time to form, there are no truly reversible processes in nature [36-38]. The second class are the irreversible processes, processes during which the system and its surroundings are out of equilibrium.

Even though Clausius laid the foundations of thermodynamics, he was still facing an important question: why are machines inefficient? He found that the solution to the problem has been anticipated 50 years before and is a thermodynamic property both fascinating and mysterious. The first glimpse of the new property was seen in 1803 by Lazarus Carnot, who in his paper "*Fundamental Principles of Equilibrium and Movement*", wrote that "*in any machine the accelerations and shocks of the moving parts represent losses of moment of activity*" [26]. In other words, any natural process has an inherent tendency to dissipate energy in unproductive ways. Lazarus' son, Sadi Carnot lifted more of the fog and almost revealed the unusual property. In 1824, he published "*Reflections on the Motive Power of Fire*" [5], which introduced the concept of heat-engines, producing work or motive from the flow of heat. Sadi Carnot was called by Clausius *"the most important of the researchers in the theory of heat"* [87]. Clausius followed the lead of the Carnots, applying his newly-laid thermodynamic concepts and following it to its root - entropy. What the Carnots found was an early statement of the second law of thermodynamics, also known as the "*entropy law*" [5]. Clausius' most important paper, *On the Moving Force of Heat* [1] (1850), first stated the basic ideas of entropy and the second law of thermodynamics. He thus answered the question of the inefficiency of heat engines. Clausius wrote that he "*intentionally chose the word entropy as similar as possible to the word energy*" [3], thus implying the close relation between entropy and energy. Like energy, "*entropy*" is a state function. It represents a measure of energy that cannot be converted into work by a closed system, as opposed to "*free energy*" that can be converted into work [66]. Entropy was introduced with the following summary sentences stating the first and second laws of thermodynamics as "*The energy of the universe is constant; the entropy of the universe tends to a maximum.*" [1]. Thermodynamic entropy as a property of a material is a consequence of

1) Translation, rotation, vibration, electron states, magnetic states, nuclear states, etc.
2) Arrangement of the particles in a crystal lattice.

Thermodynamic entropy is calculated as

$$S = S_0 + \int_{T=0}^{\tau} \frac{C_p}{T} dT \qquad (1)$$

$C_p$ is heat capacity at constant pressure, $T$ is temperature, and $S_0$ is the zero-point or residual entropy of the material at absolute zero. Note that the thermodynamic entropy equation contains two conceptually different types of entropy:

a) *Thermal entropy*, $S_{Therm} = \int (C_p / T) \, dT$, due to thermally induced motion of particles, and
b) *Residual entropy*, $S_0$ due to the arrangement of the particles in a crystal lattice.

To obtain the absolute value of entropy, we use the *Third law of thermodynamics* that states that thermodynamic entropy equals zero at absolute zero for a perfect crystal: $S_0(perfect\ crystal) = 0$.

Energy is a conserved property. Thermodynamic entropy is also a conserved property in a reversible process [22, 23]. Entropy is a non-conserved property in an irreversible process [29]. Entropy for the first time gave a quantitative difference between reversible and irreversible processes. A reversible process generates no entropy [35]. Since all natural processes are irreversible, the entropy of the universe always increases [35-38]. The thermodynamic entropy is sometimes called "*the arrow of time*" [105, 106].

Classical thermodynamics is a macroscopic theory of matter. A macroscopic system is composed of very large numbers of constituents (i.e. atoms, molecules). The state of the system is described by the average thermodynamic properties of those constituents. Thermal entropy, $S_{Therm}$, is defined in classical thermodynamics as a state function,

$$dS_{Therm} = \frac{dQ_{rev}}{T} \qquad (2)$$

$Q_{rev}$ is heat exchanged in a reversible process, $T$ is temperature, and $S_{Therm}$ is in J/K. Thus, more energy in form of heat at the temperature $T$ implies greater entropy, and $T\, dS_{Therm}$ is a measure of thermal energy at a given temperature. In any process where the system gives up energy $\Delta E$, and its entropy falls by $\Delta S_{Therm}$, a quantity at least $Q = T\, \Delta S_{Therm}$ of that energy must be given up to the system's surroundings as heat. From a macroscopic perspective, in classical thermodynamics, entropy is interpreted as a state function of a thermodynamic system: that is, a property depending only on the current state of the system. Thermal entropy $S_{Therm}$ and residual entropy $S_0$ should be considered as non-negative properties ($S_{Therm} \geq 0$; $S_0 \geq 0$). Thus, it seems that Schrödinger's negentropy concept does not have any physical sense. Total entropy $S$ is defined to be zero for an ideal crystal at absolute zero by the third law. Entropy can be determined directly by calorimetry, using equation (1). Alternatively, entropy can be measured indirectly from Gibbs free energy $\Delta G$ and enthalpy $\Delta H$, using the Gibbs equation:

$$\Delta S = (\Delta H - \Delta G)/T \qquad (3)$$

Gibbs energy is a thermodynamic property that measures the maximum amount of non-expansion or reversible work that may be performed by a thermodynamic system at a constant temperature and pressure. The Gibbs energy, originally called "*available energy*" was described as: "*the greatest amount of mechanical work which can be obtained from a given quantity of a certain substance in an initial state, without increasing its total volume or allowing heat to pass to or from external bodies, except such as at the close of the processes are left in their initial condition*" [40]. Gibbs assumed two types of energy: available energy that can be converted into a work, and unavailable energy that cannot be converted into work. Entropy is a measure of unavailable energy. "*Free energy - the available energy, energy capable of doing work, and the correct significance of the second law of thermodynamics is that free energy cannot be recycled fully because real processes are not reversible*" [82].

### 2.2. Nonequilibrium thermodynamics

The entropy equations discussed above hold for closed and isolated thermodynamic systems. Prigogine extended the use of entropy to open thermodynamic systems. Prigogine's equation for total entropy change of an open system is

$$dS = d_e S - d_i S \qquad (4)$$

$d_eS$ denoting entropy exchange with the surroundings, $d_iS$ the production of entropy due to irreversible processes in the system, like chemical reactions, diffusion, and heat transport. The term $d_iS$ is always positive, according to the second law; $d_eS$, however, may be negative as well as positive [60, 61]. To determine the value of $d_iS$, one needs to consider the ways in which an open system can exchange energy with its surroundings: work, heat and mass flow. Exchange of energy as work does not lead to entropy exchange [12]. Heat flow does lead to entropy exchange, the relationship being

$$d_eS_Q = \int_\Sigma d\left(\frac{Q}{T_b}\right)$$

where $d_eS_Q$ is entropy exchange due to heat transfer, $Q$ is the amount of heat exchanged, $T_b$ is the temperature at the boundary between a system and its surroundings and the integral is over the boundary surface area $\Sigma$ [12]. In case heat flow and $T_b$ are constant during the process over the entire surface area, the equation simplifies to

$$d_eS_Q = \frac{Q}{T_b}$$

Mass flow can also lead to entropy exchange. Entropy is a fundamental property of all substances. Substances that enter the system brings their entropy in, while those exiting takes their entropy out of the system

$$d_eS_m = \sum_{in} ms - \sum_{out} ms$$

where $d_eS_m$ is entropy exchange due to mass transfer, $m$ is mass exchanged and $s$ is specific entropy (entropy per unit mass) [12]. The first summation is over all matter entering the system, while the second summation is over all matter exiting the system. The entropy exchange with the surroundings is the sum of the heat and mass flow contributions: $d_iS = d_iS_Q + d_iS_m$ [12].

While the Clausius inequality tells us that entropy change in a process is greater or equal to heat divided by temperature in a closed system, the Prigogine equation allows us do extent thermodynamics to open systems and to find out exactly how much it is greater than $Q/T$. This difference between entropy and $Q/T$ is taken into account by the entropy production term $d_iS$ [119]. Entropy production of a thermodynamic system can be estimated through the equation

$$d_iS = \int_t \int_V \sigma \, dV dt$$

where the integral is over the volume of a system and the time period of interest. The parameter $\sigma$ is the entropy production rate, also known as the entropy source strength [12, 119]. Entropy production rate $\sigma$ is defined as a product of conjugate thermodynamic flows $J$ and forces $X$:

$$\sigma = \sum_i J_i X_i$$

where the summation is over all flows and forces acting on the system [119]. Each thermodynamic force $X_i$ acting on a system has its conjugate flow $J_i$. Examples of processes that lead to entropy production are irreversible heat transfer, mass transfer, viscous dissipation and chemical reactions [119].

Nonequilibrium thermodynamics can be divided into two regions: near-equilibrium linear region and far-from-equilibrium nonlinear region [119, 120]. In the linear region, the generalized flows and forces are related by a set of simple equations, called phenomenological equations

$$J_i = \sum_j L_{ij} X_j$$

where $L_{ij}$ is a phenomenological coefficient of force $j$ on flow $i$. Notice that a flow can be influenced by forces other than its conjugate force. For example, irreversible heat transfer can create mass diffusion [119]. Phenomenological equations are very useful because once the phenomenological coefficients are known: all one needs to find entropy production in the linear region are generalized forces, the flows being set as their functions by phenomenological equations.

Nonequilibrium thermodynamics is of particular importance in life sciences [118, 120]. Living organisms are open systems far from equilibrium [118, 120, 121]. Therefore, analyzing biological processes requires nonequilibrium thermodynamics [118, 120]. Balmer [12] gives a good introduction into nonequilibrium thermodynamics, along with its applications. For a detailed description of nonequilibrium thermodynamics, see Demirel [119].

## 2.3. Boltzmann and Statistical Entropy

Statistical mechanics describes the behavior of thermodynamic systems starting from the behavior of their constituent particles. To describe the motion of a particle, we must know three coordinate parameters plus time. The parameters needed to completely describe a physical system are the degrees of freedom. Since one mole of monoatomic ideal gas is made of $6 \cdot 10^{23}$ particles, it has $18 \cdot 10^{23}$ degrees of freedom that we have to know. Obviously, dealing with each particle individually is impossible in practice and statistics is used to simplify the problem through the concepts of microstates and their probabilities. Imagine a system made of an ideal gas at equilibrium with its many atoms moving randomly. Now take a snapshot of that system in a single instant. You just determined the system's microstate at that point in time. Since the particles move, a snapshot of the gas made at some other moment will differ. The microstate changes in time, even though the gas is at macroscopic equilibrium. So, many microstates constitute one macrostate. While only two parameters are required to describe the macrostate, describing a microstate requires describing each degree of freedom, typically of the order of $10^{23}$. A way out is offered by the Ergodic theorem, which says that the macroscopic properties of a system can be found as probability weighed average of the values for microstates [89]. This rule is also known as the Gibbs postulate [90] or second postulate of statistical thermodynamics [89]. Internal energy is defined

$$U = \sum_i p_i \varepsilon_i \qquad (5)$$

where $p_i$ is the probability of microstate i and $\varepsilon_i$ is its energy [89-91]. Therefore, we don't need to know all $10^{23}$ degrees of freedom for the microstates to find the internal energy of the system; we just need to know their probabilities and energies. The probability of a microstate as a function of its energy is given by the Boltzmann distribution

$$p_i = \frac{e^{-\varepsilon_i/k_B T}}{\sum_i e^{-\varepsilon_i/k_B T}} \qquad (6)$$

where $T$ is temperature and $k_B$ is the Boltzmann constant. The Gibbs entropy equation allows us to calculate the entropy of a system based on the probability distribution of the microstates

$$S = -k_B \sum_i p_i \ln p_i \qquad (7)$$

Consider a system in a macrostate which corresponds to a set of microstates of equal energy, volume and number of particles. Since there is nothing that would make any single microstate more probable than the others, each microstate is equally probable [74, 89, 90, 93]. This is known as the Laplace principle [93, 74], principle of equal *a priori* probabilities [90] or the first postulate of statistical thermodynamics [89]. Because of this and the fact that probabilities of all states add up to 1, for all states, $p_i = 1/\Omega$, where $\Omega$ is the total number of microstates [89, 92]. When we substitute this into the Gibbs entropy equation and sum over all $\Omega$ possible states [89, 92], it leads to

$$S = -k_B \sum_{i=1}^{\Omega} p_i \ln p_i = -k_B \sum_{i=1}^{\Omega} \frac{1}{\Omega} \ln \frac{1}{\Omega} = -k_B \cdot \Omega \cdot \left(\frac{1}{\Omega} \ln \frac{1}{\Omega}\right) = k_B \ln \Omega$$

which is the Boltzmann equation. The Boltzmann equation, also known as the Boltzmann-Planck equation, in its more well-known form is

$$S = k_B \ln W \qquad (8)$$

where $W$ is the number of microstates available to the system [91, 94]. Sometimes the microstates of the system don't all correspond to the same macrostate. In that case if we drew all microstates on a 2D graph it would resemble a map. Different territories (e.g. countries) on the map would correspond to different macrostates. The macrostates (countries) divide between them the microstates (land), but not equally; some territories are larger than others [95]. All microstates are still equally probable, but because some macrostates contain more microstates than others, they are more probable. The Boltzmann equation can still be applied to this problem, but now we apply it to each macrostate. The entropy of a macrostate is

$$S = k_B \ln W_i$$

$W_i$ is the number of microstates corresponding to that macrostate [95]. The total number of microstates available to the system is equal to the sum of $W_i$'s for all the macrostates. The number of microstates corresponding to the equilibrium macrostate $W_{eq}$ is much larger than that of all other macrostates, or in other words the equilibrium macrostate territory is by far the largest on the map, so we can say that its logarithm is approximately equal to the logarithm of the total number of microstates $W$ [91, 95]

$$S_{eq} = k_B \ln W_{eq} \approx k_B \ln W$$

Note that it is the logarithms of $W$ and $W_{eq}$ that are approximately the same, while their values may differ greatly [91, 95]. This equation enabled Boltzmann to define entropy for both equilibrium and nonequilibrium systems [95]. Alternative ways to derive the Boltzmann equation are given in [90, 91, 94]. The Gibbs and Boltzmann entropy equations allow us to understand entropy at the level of particles. The Boltzmann equation tells us that entropy is proportional to the number of microstates available to the system. Boltzmann viewed entropy as a measure of disorder. In particular, it was his attempt to reduce the second law to a stochastic collision function, or a law of probability following from the random mechanical collisions of particles. Particles, for Boltzmann, were gas molecules colliding like billiard balls in a box. Each collision brings more disorder to the nonequilibrium velocity distributions, leading to a final state of macroscopic uniformity and maximum microscopic disorder - the state of maximum entropy. Maximum entropy and the second law, he argued, are simply the result of the fact that in a world of mechanically colliding particles disordered states are the most probable [7]. Because there are so many more possible disordered states than ordered ones, a system will almost always be found either moving towards or being in the state of maximum disorder – the macrostate with the greatest number of accessible microstates, such as a gas in a box at equilibrium. On the other hand, a dynamically ordered state, one with molecules moving "*at the same speed and in the same direction*", Boltzmann concluded, is thus "*the most improbable case conceivable...an infinitely improbable configuration of energy*"[7]. Thus, entropy represents the disorder of a thermodynamic system.

### 2.4. Schrödinger: Living organisms and Negentropy

Based on Boltzmann's reasoning that entropy is a measure of disorder, Schrödinger introduced a quantitative measure of order – negentropy. Negentropy was proposed as entropy taken with a negative sign (-(entropy)). The second law of thermodynamics is the "entropy law" and represents a law of disorder, a view due to Boltzmann: "*Because there are so many more possible disordered states than ordered ones, a system will almost always be found either in the state of maximum disorder*" [7]. Disorder and order in living organisms were considered by Schrödinger, who argued: "*Life seems to be orderly and lawful behavior of matter, not based exclusively on its tendency to go over from order to disorder, but based partly on existing order that is kept up… If D is a measure of disorder, its reciprocal, 1/D, can be regarded as a direct measure of order. Since the logarithm of 1/D is just minus the logarithm of D, we can write Boltzmann's equation thus:*

$$-(entropy) = k \log(1/D)$$

(where $k$ is the Boltzmann constant $k_B$, and *log* is the natural logarithm). *Hence the awkward expression negative entropy can be replaced by a better one: entropy, taken with the negative sign, is itself a measure* *of order*" [6]. Schrödinger postulated a local decrease of entropy for living systems, quantifying it with negentropy and explaining it through order of biological structures. Order (1/*D*) represents the number of states that cannot arrange randomly, exemplified by replication of genetic code. Negentropy has found an application in biothermodynamics - thermodynamics of living organisms, describing their orderliness and explaining the general patterns of metabolism. However, recent studies are casting doubt on the validity of negentropy [46-49].

What exactly is the "*negentropy concept*" or more precisely - entropy taken with a negative sign? Boltzmann predicted in the 19th century that organisms decrease their entropy during life: "*The general struggle for existence of animate beings is not a struggle for raw materials, but a struggle for* (negative) *entropy, which becomes available through the transition of energy from the hot sun to the cold earth*" [7]. Schrödinger followed his lead: "(An organism) *feeds upon negative entropy, attracting, as it were, a stream of negative entropy upon itself, to compensate the entropy increase it produces by living and thus to maintain itself on a stationary and fairly low entropy level*" [6]. Thus, thermodynamic entropy of a cell or an organism is predicted to decrease during its life span, paralleled by an accumulation of information [12]. Balmer argued in 2006 that: "*one characteristic that seems to make a living system unique is its peculiar affinity for self-organization*" [12]. As the system lives, it grows and ages and generally becomes more complex. So, "*living systems are uniquely characterized by decreasing their entropy over their life spans*" [12]. However, Schrödinger pointed out a potential conflict between the second law and life processes, because direction of change in entropy tends to its maximum, and the direction of change in life process seemed to be toward greater order, decreasing thermodynamic entropy and accumulating information [6]. Schrödinger explained this "apparent" contradiction by suggesting that the very existence of living systems depends on increasing the entropy of their surroundings [13]. The second law is not violated but only locally circumvented at the expense of global increase in thermodynamic entropy, concluded Morowitz [13]. It seems obvious that there is at least an apparent contradiction between the second law and life processes. The decrease of thermodynamic entropy of living systems during their life spans was predicted by Boltzmann and Schrödinger and is represented by the following equation of entropy rate balance on a living system [12]

$$\frac{\dot{Q}}{T_b} + \sum_{in} \dot{m}s - \sum_{out} \dot{m}s + \frac{d_i S}{dt} = \frac{dS}{dt} \qquad (10)$$

where $\dot{Q} = dQ/dt$ is metabolic heat transfer rate, $\dot{m}$ is the mass flow, $d_i S$ is the system's entropy production, $S$ is the total system entropy and $t$ is time. This equation is interpreted by noting that $\dot{Q}$ is always negative for living organisms, according to Balmer [12]. Balmer [12] also interprets that "*since the entropy of the incoming food is lower than the entropy of the outgoing wastes, whole term $\Sigma_{in} \dot{m}s - \Sigma_{out} \dot{m}s$ is negative*." The last term $\dot{S}_P$ must always be positive, according to the second law [12]. So Balmer [12] concluded

$$\left| \frac{\dot{Q}}{T_b} + \sum_{in} \dot{m}s - \sum_{out} \dot{m}s \right| > \dot{S}_P$$

In that case for a living system:

$$\frac{dS}{dt} = m\frac{ds}{dt} + s\frac{dm}{dt} < 0$$

"*Organisms continuously decrease their thermodynamic entropy over their life spans*" [12], implying negentropy. The negentropy concept is also supported by Mahulikar and Herwig [8], Davis and Rieper [9], Ho [10. 11] and others [97, 100-102].

However, several recent studies are revealing an increase of entropy during life processes, thus questioning the negentropy concept. Hayflick [43-45] revealed an increase in entropy during the aging process. Silva [46] found an increase of entropy during lifespan of average human individuals to be 11.4 kJ/K per kg of body mass. Gems and Doonan [42] reported that the entropy of the *C. elegans* pharynx tissues increases as the animal ages, showing that Silva's [46] results are a general pattern among living organisms. Cells and organisms grow as a consequence of internal accumulation of matter, a process shown to increase their entropy [49, 118]. Thermodynamic entropy increases proportionally to mass of an organism during growth [118]. Hansen concluded that entropy of an organism "*doesn't have to decrease*" [47, 48]. Theoretical considerations indicated possible contradictions involving the negentropy concept. Instead of negative entropy change ($\Delta S<0$) Schrödinger introduced entropy taken with a negative sign ($S<0$) [6]. However, entropy itself cannot have a negative sign according to the third law of thermodynamics [12, 90, 91, 94]. Thermodynamic equation for negentropy results from a mathematically correct manipulation of the Boltzmann equation [14]. However, it has no physical sense, since entropy (just as energy) cannot have a negative value. [14]. Entropy is a thermodynamic property, resulting from motion of the particles in a system. Motion of particles cannot be negative. Therefore, thermodynamic entropy represents a non-negative property. However, entropy change ($\Delta S$) can be negative. Negentropy in the discussion above was interpreted using nonequilibrium thermodynamics, a discipline developed and applied to living organisms by Prigogine [60, 61]. Nonequilibrium thermodynamics was applied to living organisms by Garbi and Larsen [99], yielding a detailed analysis of metabolic processes. On the other hand, Gladishev [98] proposes an alternative method of analysis of living organisms.

### 2.5. Giauque and Residual Entropy

Residual entropy ($S_0$) was introduced in the first half of $20^{th}$ century and is a property of a thermodynamic system at absolute zero, appearing as a consequence of random arrangement of asymmetric atoms or molecules aligned in a crystal lattice [14, 20-21, 31-34]. Molecules can be symmetric (Figure 1a) or asymmetric (Figure 1b). They can be unaligned, like in gasses, or aligned in arrays such as crystals. Arrays of aligned molecules can be monotonic or nonmonotonic. In a monotonic array all molecules are aligned in a same way (Figure 1d). A nonmonotonic array is made of molecules that are not aligned in a same way (Figure 1c). Asymmetrical molecules can align nonmonotonically to form an imperfect crystal or can align monotonically to form a perfect crystal, an ability that results in residual entropy. Residual entropy represents a difference in entropy between an imperfect crystal and an equilibrium state - a perfect crystal. Thus, the total entropy of an imperfect crystal at absolute zero is equal to its residual entropy according to equation (1). Notice that both perfect and imperfect crystals appear at the same absolute zero temperature. The residual entropy occurs if a material can exist in different states when cooled to absolute zero. Residual entropy is a consequence of molecular arrangement in a crystal lattice and does not result from any form of molecular motion, including the "zero-point energy" of vibration or rotation [21]. The "zero-point energy state" is the same for both perfect and imperfect crystals [21]. The *"Residual entropy (3–12 J mol$^{-1}$ K$^{-1}$) is present near 0 K in some crystals composed of asymmetric atoms and molecules, for example, H, CO, N$_2$O, FClO$_3$, and H$_2$O."* [20]. Water ice is one of the first discovered examples of residual entropy, first pointed out by Pauling [25].

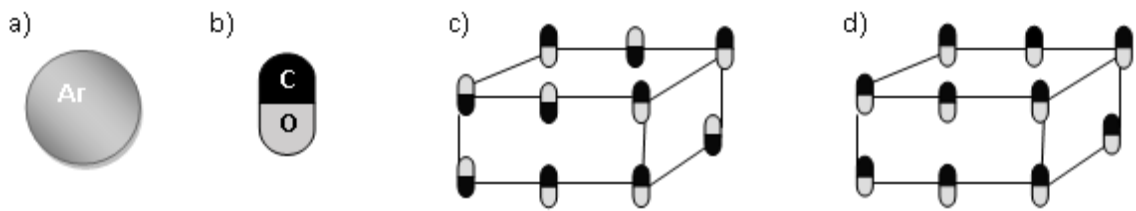

**Figure 1:** (a) A symmetric particle: in this case an argon atom, which can have only one orientation in a crystal lattice. (b) An asymmetric particle: in this case CO, which can have two orientations in a crystal lattice: either C or O can point upwards. (c) An imperfect crystal of CO: each CO molecule in the crystal lattice can point either up or down. Therefore, there is randomness and thus uncertainty in molecular orientation, leading to residual entropy. (d) A perfect crystal of CO: all molecules in the lattice are oriented in the same direction. There is no randomness nor uncertainty and, thus, residual entropy is zero.

Residual entropy is experimentally determined as the difference in entropy between an imperfect and a perfect crystal, using calorimetry. However, there's a catch: calorimetry can measure only the change in entropy by adding known amounts of heat. The situation is similar to a problem where one has two vessels, the first is empty and the second is partly filled with water. The task is to determine how much water is in the second vessel only by adding known amounts water. The solution is to fill both vessels completely and subtract the amounts of water added to them. Similarly, one adds known amounts of heat until both the perfect and the imperfect crystals melt. Since in the liquid there is no residual entropy [109-112], the entropies of both samples are equal. However, since the entropy of the starting crystals were not equal, the heat added to them was not equal either. The difference in the added heat is then converted into difference in entropy, yielding the residual entropy. An example of such a measurement is the determination of residual entropy of glycerol by Gibson and Giauque [34], who measured the heat capacity difference between a perfect and an imperfect crystal of glycerol. In the case of glycerol [34] there were both a perfect and an imperfect crystal, so the residual entropy was found directly as the difference in their experimental entropies. On the other hand, in the case of CO only an imperfect crystal was available [31]. In order to circumvent this problem, the entropy measurement for the hypothetical perfect crystal of CO was replaced by a statistical mechanical calculation based on spectroscopic measurements [31]. The calculation yielded the entropy change from a perfect crystal of CO at absolute zero to gaseous CO at its boiling point. The entropy change from the available imperfect crystal was determined by calorimetric measurements of heat capacity and enthalpies of phase changes from 0 K to the temperature of the gas [31]. The difference between the measured imperfect crystal and calculated perfect crystal values gave the residual entropy of 4.602 J mol$^{-1}$ K$^{-1}$ [31]. In terms of the water vessels, here one is given only the partially full vessel and the dimensions of the empty vessel, from which he calculates its volume. Residual entropy can also be calculated from theory, without any experiments. Kozliak [20, 33] described four ways to calculate residual entropy. The informational or "*combinatoric method*", derived using the "*coin tossing model*", is traditionally used in textbooks to illustrate residual entropy [20, 72]. Residual entropy is the difference in entropy between a non-equilibrium imperfect crystal state and the equilibrium perfect crystal state of a substance. The entropy difference between the two states can be found by applying the Boltzmann–Planck equation:

$$S_0 = k_B \ln\left[\frac{W_{2,random}}{W_{1,perfect}}\right] \quad (9)$$

where $W_2$ and $W_1$ correspond to the numbers of microstates of the imperfect and perfect crystal state, respectively [20]. The perfect crystal state has only one microstate corresponding to it – all molecules are aligned monotonically in the same way, so $W_1 = 1$. The quantity $W_2$ is related to the number of distinct orientations a molecule can have in an imperfect crystal, $m$, and the number of molecules that form the crystal, $N$, by the relation $W_2 = m^N$. This way to find $W_2$ is equivalent to tossing a coin or an $m$-sided die $N$ times, thus the name "*coin tossing model*." The residual entropy of imperfect crystalline CO calculated using equation (9) is 5.76 J mol$^{-1}$ K$^{-1}$, slightly higher than the experimentally determined 4.602 J mol$^{-1}$ K$^{-1}$ [31]. To analyze entropy-information relationship, a few relevant properties in information theory should be underlined.

### 2.6. Shannon and Information Entropy

Shannon introduced in his famous paper: "A mathematical theory of communication" the basis of information theory. Shannon introduced two main properties in information theory: amount of information and Shannon information entropy [41]. First, the amount of information, or just information, given as

$$I = K \ln(M) \quad (11)$$

where $I$ is the total information of a message (string), $K$ is a constant and $M$ is the number of possible messages in a finite set from which the message came. The constant $K$ is used to convert from one base of the logarithm to another, which corresponds to switching between units of information (i.e. bit, trit, nat…). Thus we can use any logarithm base and obtain any information units as long as we choose the right value of $K$. The following example illustrates how $I$ can be found from equation (11): "*A device with two stable positions, such as a relay or a flip-flop circuit, can store one bit of information. N such devices can store N bits, since the total number of possible states is $2^N$ and $\log_2 2^N = N$.*" [41]. So, here we have $M=2^N$ possible messages that can be stored in $N$ flip-flop circuits. Thus, the set contains $M=2^N$ possible messages. Therefore, the total information of a message that comes from that set is $I = \log_2(M) = N$ bits of information. The second property, entitled information (Shannon) entropy, is introduced in the same article. Shannon entropy $\check{S}$ is the average information per symbol in a message and is defined as

$$\check{S} = -K \sum_{i=1}^{n} p_i \ln(p_i) \quad (13)$$

where $p_i$ is the probability of symbol i. Shannon entropy $Š$ and amount of information $I$ are related by the equation: $I = N \cdot Š$, where $N$ is the number of symbols in a message.

Since Shannon's work, scientists have been discussing the relationship between thermodynamic entropy, information entropy and information. Brillouin [97] concluded that information is equal to negentropy, using this relationship to resolve the Maxwell demon problem. Based on Brillouin's work, Landauer [113] found that irreversible logical operations, such as erasing data, lead to physical irreversibility, requiring a minimal generation of heat equal to $k_B T$, where $T$ is the temperature at which the operation is performed. This result is known as the Landauer principle [115]. On the other hand, Jaynes [74] found that information $I$ is equal to thermodynamic entropy $S$. Thus both thermodynamic entropy and information can be measured in the same units of bits [74]. Thermodynamic entropy $S$ expressed in bits, instead of J/K, will be denoted as $H$ in further considerations. The quantitative relationship between information and entropy was considered by Layzer [73, 114]. According to Layzer [73, 114], a system has a finite information content – information necessary for its complete description, equal to the maximum information that can be known about the system $I_{max}$. This information is divided into two main categories: known information and unknown information. Known information will be denoted $I$. Unknown information is equal to thermodynamic entropy measured in bits, $H$. Thus, thermodynamic entropy and information are related by a simple conservation law, which states that the sum of the amount of information and the entropy is constant and equal to the system's maximum attainable information or entropy under given conditions [62-65, 73]:

$$H + I = const. = H_{max} = I_{max} \qquad (14)$$

where $H$ is entropy $I$ is information, and $H_{max}$ and $I_{max}$ are the maximum possible values. Thus: "*a gain of information is always compensated for by an equal loss of entropy*" [73]. A good example is reading a book. A book contains a certain amount of information $I_{max}$, independent of how much of it is known. As one reads the book, the amount of known information $I$ increases, while the amount of unknown information, or entropy $H$, decreases. Once the book is read, $I$ reaches its maximum value of $I_{max}$.

## 3. Theoretical analysis

The entropy concept has a 150-years long, exciting and stormy history. Entropy represents a very abstract both philosophical and scientific concept. However, after a long time and extensive use in many scientific disciplines, entropy is still frequently misused and even abused in the literature [30,72, 79]. The actual situation is very well described by von Neumann's sentence: "*no one knows what entropy really is*" [80, 81]. The entropy concept itself is clear, but some of its interpretations lead to the unbelievable confusion and misuse [30, 79].

The first question about the entropy concept is: what is the physical meaning of entropy? To get some answers, we shall recall that Clausius wrote that he "*intentionally chose the word Entropy as similar as possible to the word Energy*" [3]. This implies the close relationship between entropy and energy. Yes indeed, but how close? Clausius stated later that "*The energy of the universe is constant; the entropy of the universe tends to a maximum*" [1]. This statement apparently implies that energy and entropy are different properties. However, are these two properties somehow related? Notice that these two statements are related to thermodynamic entropy only. Obviously, we may conclude that entropy is closely related to energy. The definition given by Kostic that entropy is "*associated with energy*" does not help us too much. The phrase "*associated with*" and "*related to*" does not imply the physical meaning of entropy, nor the real nature of the relationship between entropy and energy. Clausius originally intended to explain the reason why all energy produced in combustion engines cannot be converted into useful work [1]. This leads us to conclude that entropy represents a measure of *useless energy*. Indeed, thermodynamic entropy is given in units of J/K, implying that it represents a measure of (useless) energy (J) at a given temperature (K). Thus, thanks to Clausius, we can assume that thermodynamic entropy is a part of the total energy content, a measure of the fraction of useless energy at a given temperature. During the last 150 years, the Clausius' statement has disappeared from minds of many specialists. This useless fraction of energy results from chaotic motion (translation, rotation, vibration…) of particles that a system is consisted of. We can conclude a lot about the physical meaning of thermodynamic entropy from the Helmholtz free energy equation

$$A = U - TS \qquad (23)$$

$A$ is Helmholtz free energy, $U$ is internal energy, $T$ is temperature and $S$ is thermodynamic entropy. Thus

$$1 = \frac{A}{U} + \frac{TS}{U} \qquad (24)$$

Helmholtz free energy is equal to the maximum amount of P-V work that can be extracted from a system $W_{max}$ [94]. We can write

$$1 = \frac{W_{max}}{U} + \frac{TS}{U}$$

The first term $W_{max}/U$ (or $A/U$) represents the maximum fraction of the internal energy of the system that can be extracted as work. The term $TS/U$ represents the fraction of the internal energy of the system that is "trapped" as thermal energy and cannot be extracted as work. The more work we extract from the system, the less heat we can get because of the first law $U=Q+W$, so to the maximum work $W_{max}$ corresponds minimum heat $Q_{min}$. So, we could rewrite the Helmholtz free energy equation as

$$x_W + x_Q = 1 \qquad (25)$$

where $x_W = W_{Max}/U = A/U$ and $x_Q = T \cdot S/U$, and $S$ is in that case

$$S = \frac{U}{T} x_Q \qquad (26)$$

Therefore, the term $TS/U$ represents the minimum fraction of internal energy that is trapped as heat and cannot be converted into work. In that sense, entropy is a measure of the part of internal energy that cannot be converted into work, or in practical terms is useless energy.

The second question is: how many types of entropy exist? To get possible answers let us start from the origin of the entropy concept. Clausius introduced entropy ($S$) as a purely thermodynamic property [1-3] that concerns loss of energy. Boltzmann and Gibbs introduced statistical equations for thermodynamic entropy leading to the Boltzmann constant and Boltzmann distribution. About a century after Clausius, Giauque introduced residual entropy ($S_0$) [31]. The residual entropy is calculated using the same equation and Boltzmann constant, but the normal distribution. The statistical "coin" model is used [20, 33]. Shannon independently introduced information entropy ($Š$) using the same statistical "coin" model and a very similar equation (the only difference is in the constants $k_B$ and $K$ in the Gibbs and Shannon equations, respectively). Residual entropy does not result from thermal motion, but from arrangement of molecules aligned in a chain or crystal lattice of an imperfect crystal. Thus, thermodynamic entropy represents a measure of disorder of unaligned molecules in chaotic motion. The residual entropy represents a measure of disorder of aligned molecules in a chain or crystal lattice [21]. The third type of entropy – entropy of aligned signs in a message is related to linguistics and theory of communication. So, there are two types of entropy related to material science and one related to linguistics and information theory (Figure 2). The thermal and residual entropy are given in J/K, but information entropy is given in bits. All other forms of entropy that can be found in the literature can be related to one of these three basic types.

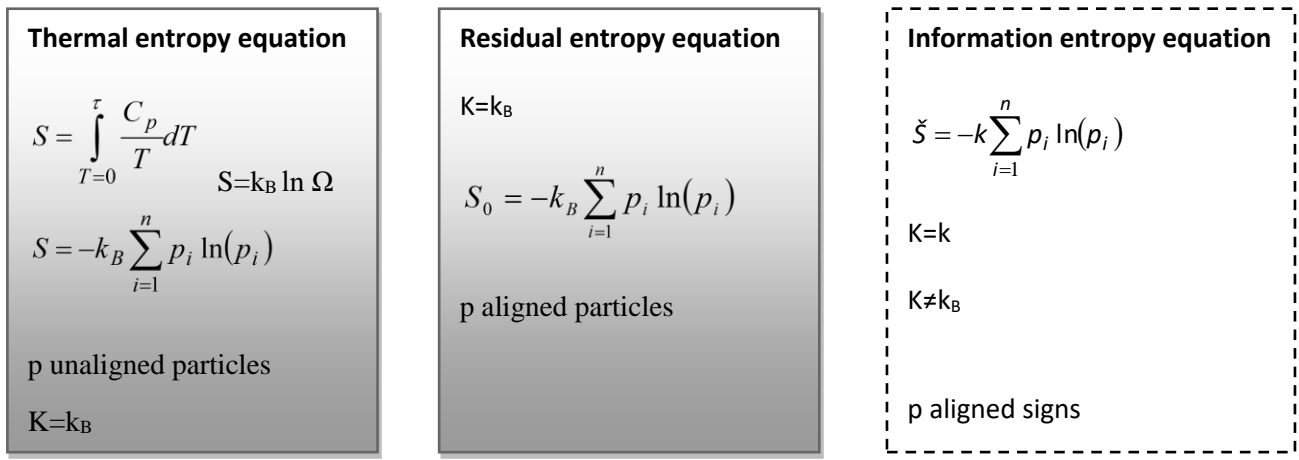

**Figure 2:** Three types of entropies: Two on the left side use Boltzmann constant and Boltzmann distribution (thermodynamic entropy) and normal distribution (residual entropy). The third is related to social sciences and information theory. $k_B$ is the Boltzmann constant and $k$ is a constant that should be determined. For thermal entropy, the probability $p_i$ is related to unaligned particles. For residual entropy the probability $p_i$ is related to aligned particles. For information entropy, $p_i$ is related to aligned symbols in a string.

Is entropy at the most fundamental philosophical – epistemological and metaphysical level a subjective or an objective parameter? Denbigh [67] stated: "*there remains at the present time a strongly entrenched view to the effect that entropy[1] is a **subjective** concept precisely because it is taken as a measure of missing information*". Jaynes' insight also suggests that the ideas of thermodynamic state and entropy are "*somewhat subjective*" [69, 74]. However, Singh and Fiorentino [70] introduced four interpretations of the entropy concept. First, "*entropy[2] as a measure of system property assumed to be **objective** parameter*" [70].

---
[1]Notice that it was not specified what kind of entropy, thermodynamic or information, or both.
[2]obviously it means the thermodynamic entropy

Second, "*entropy³ assumed as a probability for measure of information probability*" [70]. Third, "*entropy assumed as a statistic of a probability distribution for measure of information or uncertainty*" [70]. Fourth, "*entropy as a Bayesian log-likelihood functions for measure of information*" [70]. The second, third and fourth are assumed to be **subjective** parameters [70]. However, Clausius 150 years ago, and after him Gibbs, clearly stated: *"the entropy of the universe tends to a maximum"*. Indeed, thermodynamic entropy of universe tends to a maximum independently of our knowledge or even our (human) existence. To claim that thermodynamic entropy is subjective would be somewhat anthropocentric. Heat represents energy and is thus an objective parameter, and temperature represents a measure of molecular chaotic motion and thus is an objective parameter. Energy is an objective property. Thus, the measure of its useless fraction – thermodynamic entropy is also objective. Stars would explode in supernovae (and increase their entropy) independently of our ability to measure this phenomenon. Entropy change in chemical reaction occurs with or without observers. Bohm wrote "*Entropy now has a clear meaning that is independent of subjective knowledge or judgement about details of the fluctuation*" [76], and explicitly "*entropy is an objective property*" [76]. Bunge was also explicit: "*Thermodynamic probabilities are objective property*" so "*it is used to calculate another system property namely its entropy*" [77]. Hintikka wrote: *"Boltzmann and Gibbs were somewhat closer to the objective end of the spectrum of entropy as a physical objective to entropy as nonphysical property"* [78]. Further, R. Carnap wrote: "*Entropy in thermodynamics is asserted to have the same general character as temperature and heat all of which serve for quantitative characterization of some objective property of a state of a physical system*" [78]. Thus, the thermal entropy should be considered as an objective parameter. At absolute zero asymmetrical molecules of CO, $H_2O$ or H atoms would align in a crystal lattice spontaneously by a physical process, creating an imperfect crystal containing some residual entropy without an observer. Thus, the residual entropy should also be considered as an objective parameter. The experimental determination of glycerol entropy shows that residual entropy ($S_0$) is an objective parameter, since two crystals take different amounts of heat for an equal change in their temperatures [34].

Information is a subjective parameter. It depends on specific knowledge of writer and reader. It seems very easy to intuitively accept the assumption that information entropy also should be a subjective property. A thought experiment is here proposed to examine the nature of information entropy: Consider a book as a system i.e. an ancient Egyptian papyrus. It can be described by thermodynamic parameters such as volume, temperature, mass… and entropy. There are two observers – an Egyptologist and thermodynamicist. Both will measure the same mass of the system. Thus, both will notice the same thermal entropy. However, the Egyptologist recognizes and reads the hieroglyphs and can understand the message. The thermodynamicist does not understand and cannot read a message written in hieroglyphs. Thus, we may conclude that amount of information is a subjective property. However, both the thermodynamicist and the Egyptologist (using IT) can easily count the number of symbols and exactly calculate information entropy. Thus, information (Shannon) entropy represents an objective property that does not depend on subjective knowledge. It does not require understanding of a message. A message (information) is subjective, but information entropy is an objective property.

---

[3]obviously it means the Shannon entropy

Finally, let's try to answer the last question about entropy/information (thermodynamics/IT) relationship. Jaynes stated in 1957 that thermodynamic entropy can be seen as just a particular application of Shannon's information entropy to the probabilities of particular microstates of a system occurring in order to produce a particular macrostate [88, 89]. In order to resolve the Maxwell's demon paradox, Brillouin [97] concluded that negentropy is equivalent to information.

$$I = -S \qquad (27)$$

However, it seems that negentropy concept is wrong [81]. Both thermodynamic and Shannon entropy should be considered as a nonnegative property. Thermal entropy is the result of the thermal motion of particles. The thermal motion cannot be negative. Thus, entropy is a nonnegative property. Consequently, the negentropy concept is wrong. If entropy is a measure of useless fraction of total energy, and if energy is a nonnegative property, then this implies that thermal entropy is also a nonnegative property. In that case information cannot be equal to negative entropy (useless fraction of energy). Moreover, information is carried by material carriers of information aligned in a string. Thus, Information and Shannon information cannot be negative. So, entropy is nonnegative $S > 0$ and Shannon entropy is a nonnegative property $Š > 0$. Thus, Brillouine's assumption (27) is incorrect. However we know empirically that entropy and information are related. Landauer's principle [113], which was proved experimentally [122], lead us to conclude that any loss of information is followed by an increase in entropy.

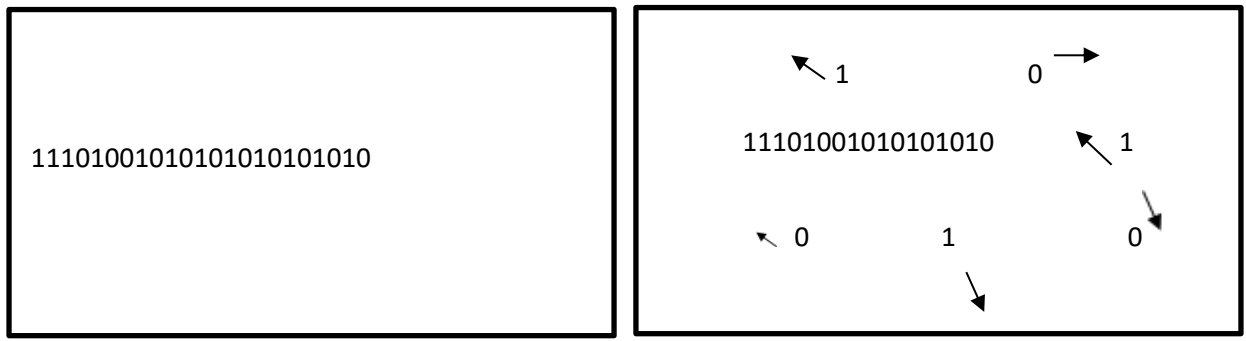

Figure 3: (left) In an imperfect crystal near absolute zero material carriers of information are aligned in an information string. Amount of information $I_0$ is at maximum. Entropy and energy is at minimum. (Right) At temperature T>0 some of the material carriers of information leaves a crystal lattice. Thus, the carried information is shortened. Unaligned particles have more degrees of freedom and more energy. Thus, entropy of the entire system increases. The total number of particles is constant during phase transition. The decrease in information (shortened string) is followed by increase in entropy. Numbers 1 and 0 represents asymmetrical particles (ie CO, OC).

Layzer suggested reciprocity of entropy and information [73]. An imperfect crystal of CO near absolute zero contains chain of aligned randomly arranged molecules or atoms (*fig 1*). Thus, it contains some residual entropy. It can be assumed that a chain of nonmonotonically aligned molecules can be understood as an information string [81]. Thus, random information contained in an imperfect crystal can be interesting for information theory. Information theory covers information contained in a string of signs and symbols and may also be interested in information carried by a crystal lattice at absolute zero. The material carriers of information aligned in an imperfect crystal leave the crystal lattice during a phase transition. Thus, the number of aligned material carriers of information decreases. The message is shortened. Consequently, the information content decreases. At the same time particles in a gas phase (ex-material carriers of information) have more degrees of freedom and energy because they perform various forms of thermal motion. Thus, thermodynamic entropy increases. Any loss of information during a phase transition from solid state imperfect crystal into gas phase is followed by an increase in thermal entropy.

If the universe consists of a single thermodynamic system and its surroundings, then an infinitesimal change in total entropy of the universe $dS_u$ is given as

$$dS_u = dS_{sys} + dS_{sur} \tag{28}$$

where $dS_{sys}$ is the infinitesimal change of entropy of the system (i.e. organism, object…) and $dS_{sur}$ is the infinitesimal change of entropy of the surroundings (rest of the universe). According to the Prigogine's equation the infinitesimal entropy change of the system is given as

$$dS_{sys} = d_e S_{sys} + d_i S_{sys} \tag{29}$$

where $d_e S_{sys}$ and $d_i S_{sys}$ are the entropy change of the system due to exchange with the surroundings and internal generation, respectively. Prigogine's equation for the surroundings is

$$dS_{sur} = d_e S_{sur} + d_i S_{sur} \tag{30}$$

where $d_e S_{sur}$ and $d_i S_{sur}$ are the entropy change of the surroundings due to exchange with the system and internal generation, respectively. By inserting equations (29) and (30) into (28) we find

$$dS_u = (d_e S_{sys} + d_i S_{sys}) + (d_e S_{sur} + d_i S_{sur})$$

$$dS_u = (d_e S_{sys} + d_e S_{sur}) + (d_i S_{sys} + d_i S_{sur}) \tag{31}$$

All the entropy generation inside the universe is taken into account by the internal entropy generation $d_i S_{sys}$ and $d_i S_{sur}$ terms. The exchange terms, $d_e S_{sys}$ and $d_e S_{sur}$, do not contribute to entropy generation in the universe. Therefore, since the universe consists solely of the system and the surroundings, the entropy imported into the system from the surroundings is equal to the entropy exported by the surroundings to the system, and vice versa. Therefore, their entropy exchanges are related as $d_e S_{sys} = - d_e S_{sur}$. Thus, equation (31) simplifies into

$$dS_u = [d_e S_{sys} + (-d_e S_{sys})] + (d_i S_{sys} + d_i S_{sur})$$

$$dS_u = d_i S_{sys} + d_i S_{sur} \tag{32}$$

In other words, only the entropy generation in the system and in its surroundings contribute to the entropy change of the universe, while the exchange terms cancel out.

Landauer [113] found that for deleting $N$ bits of information in a system, leads to an entropy generation of $\Delta_i S_{sys} = N\, k\, ln(2)$, where $k$ is the Boltzmann constant. Deleting $N$ bits of information implies an information change of the system $\Delta I_{sys} = -N\ bits$, where information $I$ is in bits. By combining these two equations, entropy generation inside the system is related to its change in amount of information:

$$\Delta_i S_{sys} = -k \cdot \ln(2) \cdot \Delta I_{sys} \qquad (33)$$

This equation can be written in differential form as

$$d_i S_{sys} = -k \cdot \ln(2) \cdot dI_{sys} \qquad (34)$$

The same reasoning can be applied to the surroundings, thus yielding

$$d_i S_{sur} = -k \cdot \ln(2) \cdot dI_{sur} \qquad (35)$$

Where $dI_{sur}$ is the infinitesimal change in information of the surroundings. Equations (34) and (35) relate the change in information to the internal generation of entropy for the system and the surroundings, respectively.

Substitution of equations (34) and (35) into the entropy of the universe equation (32) yields

$$dS_u = \left[-k \cdot \ln(2) \cdot dI_{sys}\right] + \left[-k \cdot \ln(2) \cdot dI_{sur}\right]$$

$$dS_u = -k \cdot \ln(2) \cdot \left[dI_{sys} + dI_{sur}\right] \qquad (36)$$

By definition, the sum of the information change of the system $dI_{sys}$ and the surroundings $dI_{sur}$ is the information change of the universe $dI_u$. Therefore, equation (36) becomes

$$dS_u = -k \cdot \ln(2) \cdot dI_u \qquad (37)$$

By rearranging equation (37) we find that

$$dS_u + k \cdot \ln(2) \cdot dI_u = 0 \qquad (38)$$

Equation (38) relates the entropy change of the entire universe $dS_u$ to its information change $dI_u$.

In order to arrive to the relationship between the absolute values of entropy and information of the universe, not their changes, equation (38) needs to be integrated. An indefinite integration, with no bounds, of equation (38) gives

$$\int \left[dS_u + k \cdot \ln(2) \cdot dI_u\right] = 0$$

$$\int dS_u + \int \left[k \cdot \ln(2) \cdot dI_u\right] = 0$$

$$\int dS_u + k \cdot \ln(2) \cdot \int dI_u = 0$$

$$S_u + k \cdot \ln(2) \cdot I_u + C = 0 \tag{39}$$

Where $C$ is the integration constant. Rearranging equation (39) and substituting $K = -C$, where $K$ is also a constant, gives

$$S_u + k \cdot \ln(2) \cdot I_u = K \tag{40}$$

Equation (40) gives the relationship between the absolute values of entropy and information in bits for the universe. It can be simplified if units of information are changed from bits to J/K. Information in bits is related to information in J/K as: $I_{bits} = I_{J/K} / k \ln(2)$. Therefore equation (40) becomes

$$S_u + I_{u,J/K} = K \tag{41}$$

For simplicity, we will drop the J/K from the subscript of $I$, and use information in J/K in the rest of the derivation. Therefore,

$$S_u + I_u = K \tag{42}$$

Equation (42) gives the relationship between entropy and information, both in J/K, for the universe. Since $K$ is a constant independent of both $S_u$ and $I_u$, the sum of entropy and information (given in J/K) of the universe is constant. Notice that the universe represents an isolated system. Thus, the increase in entropy should be followed by decrease in information. If entropy represents the measure of useless energy that cannot be converted into work, then it cannot be created out of nothing and should be compensated by a loss of information. If the universe is divided into two subunits, then the increase in entropy can be followed by a decrease in information in either the system, the rest of the universe, or both.

The value of the constant $K$ can be found by examining the two limiting cases. The first limiting case is if the information content of the universe is zero: $I_u=0$. In that case equation (42) implies that

$$K = S_u + 0 = S_{u,\max}$$

Therefore, constant $K$ is equal to the maximum possible value of entropy in the universe $S_{u,max}$. Since the universe expands, according to the second law of thermodynamics it increases its entropy. The entropy of the universe can, according to equation (42), increase until it reaches its maximum value $S_{u,max}$.

The second limiting case is if the entropy of the universe is zero $S_u=0$. Then equation (42) becomes

$$K = 0 + I_u = I_{u,\max}$$

This implies that the constant $K$ is also equal to the maximum possible information content of the universe $I_{u,max}$. Thus, when the limiting cases are considered, equation (42) can be rewritten as

$$S_u + I_u = K = S_{u,\max} = I_{u,\max} \tag{43}$$

This derivation shows that the sum of information and entropy is constant in an isolated system – the universe. A similar equation was proposed by Layzer [106, 114] based on logical considerations. This paper gives a formal derivation of the equation within the framework of thermodynamics. It also includes an analysis of the limitations of equation (43), finding that it is applicable only to isolated systems. The only true isolated system known in nature is the universe. So, equation (43) can give insight into processes of the expansion and increase of entropy of the universe.

## 4. Conclusions

Entropy concept is an unavoidable property in many scientific disciplines. In order to clarify the use of entropy, an attempt was made to answers the four questions from the abstract. The analysis showed that the answers are:

(1) There are only three types of entropy: two types of entropy in material science and one in information theory. So, entropy should be referred to one of the three types: any additional type of entropy seems to be artificial and can cause confusion.
(2) The physical nature of thermal entropy is represented by the statement: Thermal entropy represents a measure of **useless energy** stored in a system at a given temperature, resulting from thermal motion of particles. Residual entropy represents a measure of disorder of arrangement of aligned asymmetrical particles in a crystal lattice (string). The sum of residual and thermal entropy represents the total entropy content of a thermodynamic system.
(3) Entropy is a non-negative objective property. Thus, the negentropy concept seems to be wrong.
(4) Total entropy is reciprocal to information content, after a phase transition involving a solid state (equations 42, 43).

[101] Magalhães G. , (2015), Some Reflections on Life and Physics: Negentropy and Eurhythmy, *Quantum Matter*, Vol 4, Nr 3

[102] Brillouin L, Negentropy and Information in Telecommunications, *J. Appl. Phys*. 25, 595 (1954)

[103] Schlipp, P.A. (1973). *Albert Einstein: Philosopher-Scientist*. La Salle, IL.: Open Court Publishing.

[104] The free dictionary, (2015) http://www.thefreedictionary.com/Earth's+universe

[105] Weinert, F. (2005). *The scientist as philosopher: philosophical consequences of great scientific discoveries,* Springer, p. 143

[106] Layzer, David. (1975). "The Arrow of Time", *Scientific American*, 233:56-69

[107] Barett P., (2016) The Works of Charles Darwin: Vol 15: On the Origin of Species, vol 15, Routledge, Tazlor^Francis, London

[108] Christodoulos A. Floudas, Panos M. Pardalos,, (2006), Encyclopedia of Optimization, Springer Sience

[109] Peckham, G.D.; McNaught, I.J. Teaching Intermolecular Forces to First-Year Undergraduate Students. *J. Chem. Educ.*, **2012**, *89* (7), 955-957.

[110] Pinal, R. Effect of molecular symmetry on melting temperature and solubility. *Org. Biomol. Chem.*, **2004**, *2* (18), 2692-2699.

[111] Brown, R.J.C.; Brown, R.F.C. Melting point and molecular symmetry. *J. Chem. Educ.*, **2000**, *77* (6), 724-731.

[112] Mayer, J.E.; Brunauer, S.; Mayer, M.G. The Entropy of Polyatomic Molecules and the Symmetry Number. *J. Am. Chem. Soc.*, **1933**, *55* (1), 37-53.

[113] Landauer, R. Irreversibility and heat generation in the computing process. *IBM J. Res. Develop.*, **2000**, *44* (1/2), 261-269.

[114] Layzer, D. The arrow of time. *The Astrophysical Journal*, **1976**, *206*, 559-569.

[115] Reeb, D.; Wolf, M.M. An improved Landauer principle with finite-size corrections. *New Journal of Physics*, **2014**, *16*, 103011.

[116] A. Münster, *Classical Thermodynamics*, Wiley-Interscience, cop., London, 1970

[117] Popovic, Marko (2017): Explaining the entropy concept and entropy components. ChemRxiv. https://doi.org/10.26434/chemrxiv.5436193.v1. Retrieved: 16:53, Oct 25, 2017 (GMT)

[118] Popovic, M. Living organisms from Prigogine's perspective: an opportunity to introduce students to biological entropy balance. *Journal of Biological Education*, http://dx.doi.org/10.1080/00219266.2017.1357649
28